\documentclass[12pt]{JHEP3}

\newcommand{\be}{\begin{equation}} 
\newcommand{\ee}{\end{equation}} 
\newcommand{\bea}{\begin{eqnarray}} 
\newcommand{\eea}{\end{eqnarray}} 
\def\eqa{&=&} %\def\eqa{\!\!&=&\!\!} 
\def\ccr{\nonumber\\} 

\title{Higher spin fields from a worldline perspective}

\author{Fiorenzo Bastianelli $^{a,b}$,  Olindo Corradini $^{a,b}$  
and Emanuele Latini $^{a,c}$  \\ \vskip2mm 
$^{a}$ Dipartimento  di Fisica, Universit{\`a} di Bologna, 
via Irnerio 46, I-40126 Bologna, Italy \\ \vskip2mm  
$^{b}$ INFN, Sezione di Bologna, via Irnerio 46, I-40126 Bologna, 
Italy \\ \vskip2mm 
$^{c}$ INFN, Laboratori Nazionali di Frascati, C.P. 13, 
I-00044 Frascati, Italy \\
\vskip2mm 

E-mail:
\email{bastianelli@bo.infn.it}, \email{corradini@bo.infn.it},
\email{latini@lnf.infn.it}}

\abstract{Higher spin fields in four dimensions, 
and more generally conformal fields 
in arbitrary dimensions, can be described by spinning particle models 
with a gauged $SO(N)$ extended supergravity on the worldline. 
We consider here the one-loop quantization of these models by studying
the corresponding partition function on the one-dimensional torus.
After gauge fixing the supergravity multiplet, the partition function
reduces to an integral over the corresponding moduli space which is 
computed using orthogonal polynomial techniques.
We obtain a compact formula which gives the number of physical 
degrees of freedom for all $N$ in all dimensions.
As an aside we compute the  physical degrees of freedom of
the $SO(4)=SU(2)\times SU(2)$ model with only a $SU(2)$ factor gauged, 
which has attracted some interest in the literature.}


\preprint{}

\begin{document}

%%%%%%%%%%%%%%%%%%%%%%%%%%%%%%%%%%%%%%%%%%%%%%%%%%%%%%%%%%%%%%%%%%%
\section{Introduction}
\label{sec:intro}
%%%%%%%%%%%%%%%%%%%%%%%%%%%%%%%%%%%%%%%%%%%%%%%%%%%%%%%%%%%%%

The study of higher spin fields has attracted a great deal 
of attention in the search for generalizations of the known gauge theories
of fields of spin 1, 3/2 and 2. This search has proved to
be quite difficult, and several no-go theorems have been discovered
restricting the possible form of such generalizations.
Positive results have been achieved as well, the most 
notorious being perhaps the Vasiliev's interacting field equations, which 
involve an infinite number of fields with higher spin \cite{Vasiliev:1990en},
but an action principle for them is still lacking.
Also, many studies of free higher spin fields have been carried out,
trying to elucidate the problem further 
\cite{Fronsdal:1978rb,deWit:1979pe,Francia:2002aa}. 
Additional interest in higher spin fields arises from the study of the 
AdS/CFT correspondence in the limit of high AdS curvature, where string 
theory seems to 
reduce to a tensionless string model with an infinite tower 
of massless higher spin fields \cite{Sundborg:2000wp,Witten}. 
For a review on related topics and additional references 
see \cite{Bekaert:2005vh}.

We consider here a different perspective by studying the 
higher spin fields from a first quantized point of view.
It is known that spinning particles with a $SO(N)$ extended local 
supersymmetry on the worldline, constructed and analyzed in 
\cite{Gershun:1979fb,Howe:1988ft},
describe the propagation of particles of spin $N/2$ in four dimensions.
In fact, a canonical analysis produces the massless Bargmann--Wigner equations
as constraints for the physical sector of the Hilbert space, and 
these equations are known to describe massless particles of arbitrary spin
\cite{Bargmann:1948ck}.
More generally, the $SO(N)$ spinning particles are conformally invariant 
and describe all possible conformal free particles in arbitrary 
dimensions, as shown by Siegel \cite{Siegel:1988gd}.

We study here the one-loop quantization of the free spinning particles.
Our purpose is to obtain the correct measure on the moduli space of the 
supergravity multiplet on the one-dimensional torus, 
which is necessary for computing more general quantum corrections
arising when couplings to background fields are introduced.
As mentioned before, the introduction of couplings for such higher spin fields 
is a rather delicate matter. Nevertheless a positive result 
for the $SO(N)$ spinning particles has been obtained
in \cite{Kuzenko:1995mg}, where the couplings to de Sitter 
or anti de Sitter backgrounds are constructed.

In this paper we restrict ourselves to flat space, and calculate the path 
integral on the one-dimensional torus to obtain compact formulas 
which give the number of physical degrees of freedom of the spinning 
particles for all $N$ in all dimensions.
In addition, we look at the $SO(4)$ model introduced 
by Pashnev and Sorokin in \cite{Pashnev:1990cf},
where only a $SU(2)$ subgroup is gauged.
We compute the corresponding physical degrees of freedom
and resolve an ambiguity described there.
The particular cases of $N=0,1,2$ coupled to a curved target space have 
been discussed in \cite{Bastianelli:2002fv,Bastianelli:2002qw,Ba:2005vk}, 
respectively.
Some aspects of the path integral approach to the $SO(N)$ spinning particles 
have also been studied in
\cite{Papadopoulos:1989rg,Pierri:1990rp,Rivelles:1990dq,Marnelius:1993ba}.

We structure our paper as follows. In section 2 we review the 
classical action of the $SO(N)$ spinning particle
and remind the reader of the gauge invariances that must be gauge fixed.
In section 3  we describe the gauge fixing on the one-dimensional torus,
i.e. a circle, and obtain the measure on the moduli space
of the $SO(N)$ extended supergravity fields on the torus.
In section 4 we compute the integrals over the $SO(N)$ moduli space
using orthogonal polynomial techniques, and obtain 
the formulas for the number of physical degrees of freedom.
In section 5 we apply our techniques to the Pashnev--Sorokin model
and find that in $D=4$ the model has five
degrees of freedom, corresponding to a graviton and three scalars.
We present our conclusions and future perspectives in section 6.
We include in three appendices a brief discussion on the gauge fixing 
of the $SO(N)$ gauge fields on the torus, a review of the relation between
the Van der Monde determinant and orthogonal polynomials, and 
some technical details on the gauge fixing of the Pashnev--Sorokin model.

%%%%%%%%%%%%%%%%%%%%%%%%%%%%%%%%%%%%%%%%%%%%%%%%%%%%%%%%%%%%%%%%%%%
\section{Action and gauge symmetries} 
%%%%%%%%%%%%%%%%%%%%%%%%%%%%%%%%%%%%%%%%%%%%%%%%%%%%%%%%%%%%%%%%%%%

The minkowskian action for the $SO(N)$ spinning particle 
in flat target spacetime is given by 
\be 
S_M[X,G]=\int_0^1 dt \ \bigg[  
\frac{1}{2} e^{-1} (\dot x^\mu -i \chi_i \psi_i^\mu)^2 
+ \frac{i}{2} \psi_i^\mu 
(\delta_{ij} \partial_t -a_{ij} )\psi_{j\mu} \bigg ]  
\ee 
where $X=(x^\mu,\psi^\mu_i)$ collectively describes  the coordinates $x^\mu$ 
and the extra fermionic degrees of freedom $\psi^\mu_i$  of the spinning 
particle, and $G=(e, \chi_i, a_{ij})$ represents the set of gauge fields 
of the $N$-extended worldline supergravity,
containing the einbein, gravitinos and $SO(N)$ gauge fields.
The index $\mu=0,\dots,D-1$ is a spacetime index while $i,j=1,\dots,N$
are internal $SO(N)$ indices.
The gauge transformations on the supergravity multiplet $G$  are 
described by the gauge parameters $ (\xi, \epsilon_i, \alpha_{ij})$ and read
\bea 
\delta e \eqa \dot \xi + 2 i \chi_i \epsilon_i \ccr 
\delta \chi_i \eqa \dot \epsilon_i  - a_{ij}\epsilon_j  
+\alpha_{ij} \chi_j \ccr 
\delta a_{ij} \eqa \dot \alpha_{ij} +\alpha_{im} a_{mj} + 
\alpha_{jm} a_{im}  \ . 
\label{gauge-tr-m} 
\eea 
In the following 
we prefer to use euclidean conventions, and perform a  
Wick rotation to euclidean time $ t \to  -i \tau $,
accompanied by the Wick rotations of the $SO(N)$ gauge fields 
$a_{ij} \to  i a_{ij} $, just as done in \cite{Ba:2005vk}
for the $N=2$ model. We obtain the euclidean action
\be 
S[X,G] 
=\int_0^1 d\tau \ \bigg[  
\frac{1}{2} e^{-1} (\dot x^\mu - \chi_i \psi_i^\mu)^2 
+ \frac{1}{2} \psi_i^\mu 
(\delta_{ij} \partial_\tau -a_{ij} )\psi_{j\mu} \bigg ]  
\label{action:spinning}
\ee 
with the gauge symmetries on the supergravity multiplet given by 
\bea 
\delta e \eqa \dot \xi + 2 \chi_i \epsilon_i \ccr 
\delta \chi_i \eqa \dot \epsilon_i  - a_{ij}\epsilon_j  
+\alpha_{ij} \chi_j \ccr 
\delta a_{ij} \eqa \dot \alpha_{ij} +\alpha_{im} a_{mj} + 
\alpha_{jm} a_{im}   
\label{gauge-tr-e} 
\eea 
where we have also Wick rotated the gauge parameters 
$\epsilon_i \to -i \epsilon_i$, $ \xi \to  -i \xi $.
These are the gauge symmetries that will be fixed 
on the one-dimensional torus in the next section.

%%%%%%%%%%%%%%%%%%%%%%%%%%%%%%%%%%%%%%
\section{Gauge fixing on the torus} 
%%%%%%%%%%%%%%%%%%%%%%%%%%%%%%%%%%%%%%

Here we study the partition function on the one-dimensional torus $T^1$
\be 
Z\sim \int_{T^1} \frac{{\cal D} X {\cal D} G}{\rm Vol\,(Gauge)} \, 
e^{-S[X,G]}  \ .
\ee 
First we need to gauge fix the local symmetries.
We use the Faddeev-Popov method to extract the volume of the gauge group
and select a gauge which fixes completely the supergravity multiplet
up to some moduli. In particular, we specify a gauge where 
$(e, \chi_i, a_{ij})= (\beta, 0 ,\hat a_{ij})$ are constants. 
The gauge on the einbein is rather standard, and produces an integral 
over the proper time $\beta$ \cite{Henneaux:1982ma}. 
The fermions and the gravitinos are taken with antiperiodic boundary 
conditions. This implies that the gravitinos can be completely gauged away
as there are no zero modes for the differential operator that relates
the gauge parameters $\epsilon_i $ to the gravitinos,
see eq. (\ref{gauge-tr-e}). 
As for the $SO(N)$ gauge fields, the gauge conditions 
$a_{ij} = \hat a_{ij}(\theta_k)$  can be chosen to depend on a set of 
constant angles  $\theta_k$, 
with $k=1,...,r$, where $r$ is the rank of group $SO(N)$.
This is reviewed in appendix \ref{gf}. 
These angles are the moduli of the gauge fields on the torus and 
must be integrated over a fundamental region.
Thus, taking into account the ghost determinants, we find that 
the gauge fixed partition function reads as 
\bea 
Z \eqa -\frac{1}{2}  
\int_0^\infty \frac{d\beta}{\beta}\,  
\int \frac{d^Dx}{(2\pi \beta)^\frac{D}{2}}  
\ccr 
&\times&
K_N  
\bigg [ \prod_{k=1}^r \int_0^{2\pi} \frac{d\theta_k}{2\pi} \bigg ]  
\bigg ( {\rm Det}\, (\partial_\tau -\hat a_{vec})_{_{ABC}}  
\bigg)^{\frac{D}{2} -1}  
{\rm Det}' \, (\partial_\tau -\hat a_{adj})_{_{PBC}}   
\eea 
where $K_N$ is a normalization factor that implements the reduction 
to a fundamental region of moduli space and will be discussed shortly.
This formula contains the well-known proper time integral with the
appropriate measure for one-loop amplitudes,
and the spacetime volume integral with the standard free particle measure
($(2\pi \beta)^{-\frac{D}{2}}$). In addition, it contains
the integrals over the $SO(N)$ moduli $\theta_k$
and the determinants of the ghosts and of the remaining fermion
fields. In particular, the second line contains the determinants of the 
susy ghosts and of the Majorana fermions $\psi_i^\mu$ which all have 
antiperiodic boundary conditions (ABC) and transform in the vector 
representation of $SO(N)$. The last determinant instead is due to the ghosts
for the $SO(N)$ gauge symmetry. They transform in the adjoint representation 
and have periodic boundary conditions (PBC), so they have zero modes 
(corresponding to the moduli directions) which are excluded from the 
determinant (this is indicated by the prime on ${\rm Det}'$). 
The whole second line computes the number of physical   
degrees of freedom, normalized to one for a real scalar field, 
\bea 
Dof(D,N)= 
K_N  
\bigg [ \prod_{k=1}^r \int_0^{2\pi} \frac{d\theta_k}{2\pi} \bigg ]  
\bigg ( {\rm Det}\, (\partial_\tau -\hat a_{vec})_{_{ABC}}  
\bigg)^{\frac{D}{2} -1}  
{\rm Det}' \, (\partial_\tau -\hat a_{adj})_{_{PBC}}  \quad   
\eea 
In fact, for $N=0$  there are neither gravitinos nor gauge fields,  
$K_0=1$, and all other terms in the formula are absent 
\cite{Bastianelli:2002fv}, so that 
\be 
Dof(D,0)= 1  
\ee 
as it should, since the $N=0$ model describes a real scalar field
in target spacetime.
We now present separate discussions for even $N$ and  odd $N$, 
as typical for the orthogonal groups, and explicitate further 
the previous  general formula.

%%%%%%%%%%%%%%%%%%%%%%%%%%%%%%%%%%%
\subsection{Even case: $N=2r$}  
\label{sec:even} 
%%%%%%%%%%%%%%%%%%%%%%%%%%%%%%%%%%%%

To get a flavor of the general formula let us briefly review the $N=2$
case treated in \cite{Ba:2005vk}. We have a $SO(2)=U(1)$ gauge field 
$a_{ij}$ which can be gauge fixed to the constant value
\be 
\hat a_{ij}  =  
\left( \begin{array}{cc} 
0& \theta \\
-\theta & 0 
\end{array} \right) 
\label{so2} 
\ee 
where $\theta$ is an angle that corresponds to the $SO(2)$ modulus.
A fundamental region of gauge inequivalent
configurations is given by $\theta\in[0,2\pi]$ with identified boundary 
values: it corresponds to a one-dimensional torus. The factor 
$K_2=1$ because there are no further identifications
on moduli space, and the formula reads
\bea
Dof(D,2)\eqa
 \int_0^{2\pi} \frac{d\theta}{2\pi} \,
\bigg ( 
\underbrace{{\rm Det}\, (\partial_\tau -\hat a_{vec})_{_{ABC}}}_{
(2 \cos\frac{\theta}{2})^2}
\bigg)^{\frac{D}{2} -1} 
\underbrace{
{\rm Det}' \, (\partial_\tau )_{_{PBC}}}_{1} 
\ccr
\eqa
\left \{ \begin{array}{cc} 
 \frac{(D-2)!}{[(\frac{D}{2}-1)!]^2}  &   {\rm even}\ D
\\
0 &   {\rm odd}\ D
\end{array} \right . \ .
\eea
This formula correctly reproduces the number of physical
degrees of freedoms of a gauge $(\frac{D}{2}-1)$--form 
in even dimensions $D$.  Instead, for odd $D$, the above integral vanishes 
and one has no degrees of freedom left. This may be interpreted as due to the
anomalous behavior of an odd number of Majorana fermions 
under large gauge transformations \cite{Elitzur:1985xj}.
In this formula the first determinant is due to the $D$ Majorana 
fermions, responsible for a power $\frac{D}{2}$ of the first determinant,
and to the bosonic susy ghosts, 
i.e. the Faddeev--Popov determinant for local susy,
responsible for the power $-1$ of the first determinant.
This determinant
is more easily computed using the $U(1)$ basis which diagonalizes 
the gauge field in (\ref{so2}).
The second determinant is due the $SO(2)$ ghosts which of course do not couple 
to the gauge field in the abelian case.
A zero mode is present since these ghosts have 
periodic boundary conditions and is excluded from the determinant. 
This last determinant does not contribute to the $SO(2)$ modular measure. 

In the general case, the rank of $SO(N)$ is  $r=\frac{N}{2}$ for even $N$,
and by constant gauge transformations one can always put a constant field 
$ a_{ij}$ in a skew diagonal form 
\be 
\hat a_{ij}  =  
\left( \begin{array}{ccccccc} 
0& \theta_1 & 0 & 0 & \ . & 0 & 0 \\ 
-\theta_1 & 0 & 0 & 0 & \ . & 0 & 0 \\
0 & 0 & 0& \theta_2 & \ . & 0 & 0 \\ 
0 & 0 & -\theta_2 & 0 &  \ . & 0 & 0 \\
 . & . &  . & . &  \ . & . & .  \\ 
0 & 0 & 0 & 0 & \ . & 0 & \theta_r \\ 
0 & 0 & 0 & 0 & \ . & -\theta_r & 0 
\end{array} \right)  \ .
\label{ngauge1} 
\ee 
The $\theta_k$ are angles since large gauge transformations can be used
to identify $\theta_k\sim \theta_k + 2\pi n$
with integer $n$. The determinants are easily computed 
pairing up coordinates into complex variables that diagonalize
the matrix (\ref{ngauge1}). 
Then
\be 
{\rm Det}\, (\partial_\tau -\hat a_{vec})
= 
\prod_{k=1}^r  
{\rm Det}\, (\partial_\tau +i \theta_r)\, 
{\rm Det}\, (\partial_\tau -i \theta_r)
\ee 
and thus 
\be 
\bigg ( 
{\rm Det}\, (\partial_\tau -\hat a_{vec})_{_{ABC}} 
\bigg)^{\frac{D}{2} -1}   
= 
\prod_{k=1}^r  
\Big( 2 \cos\frac{\theta_k}{2} \Big)^{D-2}  \ .
\ee 
Similarly
\bea 
{\rm Det}' \, (\partial_\tau -\hat a_{adj})_{_{PBC}}   
\eqa  
\prod_{k=1}^r 
{\rm Det}' \, (\partial_\tau)   
\ccr 
&\times&
\prod_{k<l} 
{\rm Det} \, (\partial_\tau +i (\theta_k+\theta_l)) \,   
{\rm Det} \, (\partial_\tau -i (\theta_k+\theta_l))   
\ccr 
&\times&
\prod_{k<l} 
{\rm Det} \, (\partial_\tau +i (\theta_k-\theta_l))  \, 
{\rm Det} \, (\partial_\tau -i (\theta_k-\theta_l))   
\ccr
\eqa
\prod_{k<l} \Big (2 \sin \frac{\theta_k+\theta_l}{2} \Big )^2 
\Big (2 \sin \frac{\theta_k-\theta_l}{2} \Big )^2 \ . 
\eea 
Thus, with the  normalization factor $K_N= \frac{2}{2^r r!} $
one obtains the final formula
\bea 
Dof(D,N) \eqa \frac{2}{2^r r!}
\bigg [ \prod_{k=1}^r \int_0^{2\pi} \frac{d\theta_k}{2\pi} \,   
\Big (2 \cos\frac{\theta_k}{2} \Big )^{D-2}\bigg ]  \ccr 
&\times&  
\prod_{k<l} \Big (2 \sin \frac{\theta_k+\theta_l}{2} \Big )^2 
\Big (2 \sin \frac{\theta_k-\theta_l}{2} \Big )^2 \ . 
\label{12} 
\eea 
The normalization $K_N= \frac{2}{2^r r!}$ can be understood as follows. 
A factor $\frac{1}{ r!}$ is due to the fact that with a $SO(N)$ constant  
gauge transformation one can permute the angles $\theta_k$  
and there are $r$ angles in total. 
The remaining factor $\frac{2}{2^r}$ can be understood as follows. 
One could change any angle $\theta_k$ to $-\theta_k$ if parity would be 
allowed (i.e. reflections of a single coordinate) 
and this would give the factor $\frac{1}{2^r}$. 
Thus we introduce parity transformations, which is an invariance of  
(\ref{12}), by enlarging the gauge group by a $Z_2$ factor and 
obtain the group $O(N)$. 
This justifies the identification of $\theta_k$ with $-\theta_k$ and  
explains the remaining factor 2; equivalently, within $SO(N)$ 
gauge transformations one can only change signs to pairs of angles  
simultaneously. 
It is perhaps more convenient to use some trigonometric identities and
write the number of degrees of freedom as 
\bea 
Dof(D,N) \eqa  \frac{2}{2^r r!} 
\prod_{k=1}^r \int_0^{2\pi} \frac{d\theta_k}{2\pi} \,   
\Big (2 \cos\frac{\theta_k}{2} \Big )^{D-2} 
  \ccr &\times& 
\prod_{k<l} \left [ 
\Big (2 \cos\frac{\theta_k}{2} \Big )^2 
- \Big (2 \cos\frac{\theta_l}{2} \Big )^2  \right ]^2 \ . 
\label{dof-even}
\eea 

%%%%%%%%%%%%%%%%%%%%%%%
\subsection{Odd case: $N=2r+1$}
\label{sec:odd} 
%%%%%%%%%%%%%%%%%%%%%%%%%%

The case of odd $N$ describes a fermionic system in target space.  
In fact, the simplest example is for $N=1$, which gives a spin 1/2 fermion.
It  has been treated in \cite{Bastianelli:2002qw} on a general curved 
background, but there are no worldline gauge fields in this case.
For odd $N>1$ the rank of the gauge group is $r= \frac{N-1}{2}$ 
and the gauge field in the vector representation  $a_{ij}$ 
can be gauge fixed to a constant matrix of the form 
\be 
\hat a_{ij}  =  
\left( \begin{array}{cccccccc} 
0& \theta_1 & 0 & 0 &\ . & 0 & 0 & \ 0\\ 
-\theta_1 & 0 & 0 & 0 & \ . & 0 & 0 & \ 0\\
0 & 0 & 0& \theta_2 & \ . & 0 & 0 & \ 0\\ 
0 & 0 & -\theta_2 & 0 &  \ . & 0 & 0 & \ 0\\
 . & . &  . & . &  \ . & . & . & \ . \\ 
0 & 0 & 0 & 0 & \ . & 0 & \theta_r & \ 0\\ 
0 & 0 & 0 & 0 & \ . & -\theta_r & 0 & \ 0\\
0 & 0 & 0 & 0 & \ . & 0 & 0 & \ 0
\end{array} \right) \ .
\label{gauge2} 
\ee 
Then, in a way somewhat similar to the even case, one gets
\be 
{\rm Det}\, (\partial_\tau -\hat a_{vec})
= {\rm Det}\, (\partial_\tau) 
\prod_{k=1}^r  
{\rm Det}\, (\partial_\tau +i \theta_k)\, 
{\rm Det}\, (\partial_\tau -i \theta_k)
\ee 
and thus
\be 
\bigg ( 
{\rm Det}\, (\partial_\tau -\hat a_{vec})_{_{ABC}} 
\bigg)^{\frac{D}{2} -1}   
= 
2^{\frac{D}{2}-1} 
\prod_{k=1}^r  
\Big( 2 \cos\frac{\theta_k}{2} \Big)^{D-2}  \ .
\ee 
Similarly for the determinant in the adjoint representation
\bea 
{\rm Det}' \, (\partial_\tau -\hat a_{adj})_{_{PBC}}   
\eqa  
\prod_{k=1}^r 
{\rm Det}' \, (\partial_\tau) \,    
{\rm Det} \, (\partial_\tau +i \theta_k) \,       
{\rm Det} \, (\partial_\tau -i \theta_k)   
\ccr 
&\times&\!\! 
\prod_{k<l} 
{\rm Det} \, (\partial_\tau +i (\theta_k+\theta_l)) \,      
{\rm Det} \, (\partial_\tau -i (\theta_k+\theta_l))   
\ccr 
&\times&\!\! 
\prod_{k<l} 
{\rm Det} \, (\partial_\tau +i (\theta_k-\theta_l))  \,    
{\rm Det} \, (\partial_\tau -i (\theta_k-\theta_l))   
\eea 
which gives 
\bea 
{\rm Det}' \, (\partial_\tau -\hat a_{adj})_{_{PBC}}   
&=&
\prod_{k=1}^r 
\Big( 2 \sin\frac{\theta_k}{2} \Big)^2 
\ccr
&\times&
\prod_{k<l} \Big (2 \sin \frac{\theta_k+\theta_l}{2} \Big )^2 
\Big (2 \sin \frac{\theta_k-\theta_l}{2} \Big )^2 \ . 
\eea 
Thus, with a factor
\be 
K_N= \frac{1}{2^r r!} 
\ee 
one gets the formula 
\bea 
Dof(D,N) &=&
\frac{2^{\frac{D}{2}-1}}{2^r r!} \, 
 \prod_{k=1}^r \int_0^{2\pi} \frac{d\theta_k}{2\pi} \,   
\Big (2 \cos\frac{\theta_k}{2} \Big)^{D-2} 
\Big( 2 \sin\frac{\theta_k}{2} \Big)^2 
\ccr 
&\times& \prod_{k<l} \Big (2 \sin \frac{\theta_k+\theta_l}{2} \Big )^2 
\Big (2 \sin \frac{\theta_k-\theta_l}{2} \Big )^2 \ .
\eea 
In the expression for $K_N$ the factor 2 that appeared in the even case 
is now not included, since in the  gauge (\ref{gauge2})
one can always reflect the last coordinate 
to obtain a $SO(N)$ transformation that changes $\theta_k$ into $-\theta_k$.
 
For explicit computations it is perhaps more convenient to write 
the number of degrees of freedom as 
\bea 
Dof(D,N) &=&  \frac{2^{\frac{D}{2}-1}}{2^r r!} 
\prod_{k=1}^r  
\int_0^{2\pi} \frac{d\theta_k}{2\pi} \,   
\Big (2 \cos\frac{\theta_k}{2} \Big )^{D-2}
\Big( 2 \sin\frac{\theta_k}{2} \Big)^2  
\ccr 
&\times& 
\prod_{k<l} \left [ \Big (2 \cos\frac{\theta_k}{2} \Big )^2 
-\Big (2 \cos\frac{\theta_l}{2} \Big )^2 
\right ]^2 \ . 
\label{dof-odd}
\eea 

%%%%%%%%%%%%%%%%%%%%%%%%%%%%%%%%%%%%%
\section{Degrees of freedom}
\label{section:orto}
%%%%%%%%%%%%%%%%%%%%%%%%%%%%%%

We now compute explicitly the number of physical degrees of freedom for the
spinning particles propagating in arbitrary dimensions.
In the previous section we have obtained the expressions
which compute them, eqs. (\ref{dof-even}) and (\ref{dof-odd}), 
which we rewrite here for commodity
\bea 
Dof(D,2r) \eqa  \frac{2}{2^r r!}  
\prod_{k=1}^r \int_0^{2\pi} \frac{d\theta_k}{2\pi} \,    
\Big (2 \cos\frac{\theta_k}{2} \Big )^{D-2} 
\ccr  
&\times&
\prod_{1\leq k<l\leq r} \left [  
\Big (2 \cos\frac{\theta_l}{2} \Big )^2  
- \Big (2 \cos\frac{\theta_k}{2} \Big )^2  \right ]^2 
\label{eq:dofp}\\[4mm]
Dof(D,2r+1) \eqa  \frac{2^{\frac{D}{2}-1}}{2^r r!}  
\prod_{k=1}^r   
\int_0^{2\pi} \frac{d\theta_k}{2\pi} \,    
\Big (2 \cos\frac{\theta_k}{2} \Big )^{D-2}  
\Big (2 \sin\frac{\theta_k}{2} \Big )^2  
\ccr  
&\times&
\prod_{1\leq k<l\leq r} \left[ \Big (2 \cos\frac{\theta_l}{2} \Big )^2  
-\Big (2 \cos\frac{\theta_k}{2} \Big )^2  
\right]^2
\label{eq:dofd}
\eea 
with $N=2r$ and $N=2r+1$, respectively. It is obvious that $Dof(D,N)$ vanishes
for an odd number of dimensions  
\bea 
Dof(2d+1,N) =0,\qquad  \forall N>1
\eea   
as in such case the integrands are odd under the
$Z_2$ symmetry $\frac{\theta}{2}\rightarrow\pi- \frac{\theta}{2}$. 
Only for $N=0,1$ these models have a non-vanishing number of 
degrees of freedom propagating in an odd-dimensional spacetime, as
in such cases there are no constraints coming from the 
vector gauge fields. 
Also for $N=2$ these models can have degrees of freedom propagating 
in odd-dimensional target spaces, provided a suitable Chern-Simons
term is added to the worldline action~\cite{Howe:1988ft}. 
However, Chern-Simons couplings are not possible for $N>2$.
 
To compute (\ref{eq:dofp}) and (\ref{eq:dofd}) for an even-dimensional 
target space we make use of the orthogonal polynomials method reviewed 
in appendix \ref{sec:VdM}. In order to do that, we first observe that the 
integrands are even under the aforementioned $Z_2$ symmetry, and thus 
we can restrict the range of integration
\bea 
Dof(D,2r) \eqa  \frac{2}{r!}  
\prod_{k=1}^r \int_0^{\pi} \frac{d\theta_k}{2\pi} \,    
\Big (2 \cos\frac{\theta_k}{2} \Big )^{D-2}  
\ccr  
&\times&
\prod_{1\leq k<l\leq r} \left [  
\Big (2 \cos\frac{\theta_l}{2} \Big )^2  
- \Big (2 \cos\frac{\theta_k}{2} \Big )^2  \right ]^2  \ , \\[4mm] 
Dof(D,2r+1) \eqa  \frac{2^{\frac{D}{2}-1}}{r!}  
\prod_{k=1}^r   
\int_0^{\pi} \frac{d\theta_k}{2\pi} \,    
\Big (2 \cos\frac{\theta_k}{2} \Big )^{D-2}  
\Big (2 \sin\frac{\theta_k}{2} \Big )^2  
\ccr  
&\times&
\prod_{1\leq k<l\leq r} \left[ \Big (2 \cos\frac{\theta_l}{2} \Big )^2  
-\Big (2 \cos\frac{\theta_k}{2} \Big )^2  
\right]^2~. 
\eea 
Now, upon performing the transformations $x_k = \sin^2 
\frac{\theta_k}{2}$, we get 
\bea 
Dof(2d,2r) \eqa \frac{2^{2(d-1)r +(r-1)(2r-1)}}{\pi^r r!} 
\ccr  &\times&
\prod_{k=1}^r\int_0^1 dx_k\ x_k^{-1/2}(1-x_k)^{d-3/2} 
\prod_{k<l} (x_l-x_k)^2 \ , \label{even:VdM}\\[4mm] 
Dof(2d, 2r+1) \eqa \frac{2^{(d-1) +r(2r+2d-3)}}{\pi^r r!}  
\ccr  &\times&
\prod_{k=1}^r\int_0^1 dx_k\ x_k^{1/2}(1-x_k)^{d-3/2} 
\prod_{k<l} (x_l-x_k)^2 \ .
\label{odd:VdM} 
\eea 
We have made explicit in the integrands the square of the Van der Monde 
determinant: it is then possible to use the orthogonal 
polynomials method to perform the multiple integrals. 
Note in fact that in (\ref{even:VdM}) and (\ref{odd:VdM}) the 
prefactors of the Van der Monde determinant have the correct form to be 
weights $w^{(p,q)}(x) =  
x^{q-1}(1-x)^{p-q}$ for the Jacobi polynomials $G_k^{(p,q)}$ with 
$(p,q)=(d-1,1/2)$ and $(p,q)=(d,3/2)$, respectively. 
The integration domain is also 
the correct one to set the orthogonality conditions 
\bea 
\int_0^1 dx\ w(x) G_k(x) G_l(x) = h_k(p,q)\, \delta_{kl} 
\eea 
with the normalizations given by 
\bea 
h_k(p,q) = {k!\, \Gamma(k+q)\Gamma(k+p)\Gamma(k+p-q+1)\over 
  (2k+p)\Gamma^2(2k+p)} \ ,
\eea 
see~\cite{abramo} for details about the known orthogonal polynomials. 
Since the Jacobi polynomials $G_k^{(p,q)}$ are all monic, 
the even$-N$ formula reduces to 
\bea                    
Dof(2d,2r) \eqa  {2^{2(d-1)r +(r-1)(2r-1)}\over \pi^r }  
\ \prod_{k=0}^{r-1} h_k(d-1,1/2)\ccr\eqa 
2^{(r-1)(2r+2d-3)}{\Gamma(2d-1)\over\Gamma^2(d)} 
{1\over \pi^{r-1}} \prod_{k=1}^{r-1} h_k(d-1,1/2)  
\eea 
where in the second identity we have factored out the normalization of 
the zero-th order polynomial. It is straightforward algebra to get rid of 
all the irrational terms and  reach the final expression 
\bea 
Dof(2d,2r) = 2^{r-1} {(2d-2)!\over [(d-1)!]^2}\ \prod_{k=1}^{r-1} 
{k\ (2k-1)!\ (2k+2d-3)!\over 
  (2k+d-2)!\ (2k+d-1)!}~. 
\eea 
We have checked that these numbers correspond to the dimensions 
of the rectangular $SO(N)$ Young tableaux with 
$(D-2)/2$ rows and $N/2$ columns.

For odd $N$ we have instead 
\bea 
Dof(2d,2r+1) \eqa {2^{(d-1) +r(2r+2d-3)}\over \pi^r }  
\ \prod_{k=0}^{r-1} h_k(d,3/2)\ccr\eqa 
{2^{(2-d)+r(2r+2d-3)}\over d}{\Gamma(2d-1)\over\Gamma^2(d)} 
{1\over \pi^{r-1}} \prod_{k=1}^{r-1} h_k(d,3/2)  
\eea 
which can be reduced to 
\bea 
Dof(2d,2r+1)={2^{d-2+r}\over d}{(2d-2)!\over [(d-1)!]^2} 
\ \prod_{k=1}^{r-1} 
{(k+d-1)\ (2k+1)!\ (2k+2d-3)!\over 
  (2k+d-1)!\ (2k+d)!} \quad
\eea 
and again we have checked that
these numbers correspond to the dimensions 
of the spinorial rectangular $SO(N)$ Young tableaux 
with $(D-2)/2$ rows and $(N-1)/2$ columns.

From these final expressions we can single out a few interesting special cases 
\bea 
(i)\quad Dof(2,N) \eqa 1,\qquad \forall N\\[2mm] 
(ii)\quad Dof(4,N) \eqa 2,\qquad \forall N\\[2mm]   
(iii)\quad Dof(2d,2) \eqa {(2d-2)!\over [(d-1)!]^2}\\[2mm]  
(iv)\quad Dof(2d,3) \eqa {2^{d-1}\over d} {(2d-2)!\over 
  [(d-1)!]^2}\\[2mm]  
(v)\quad Dof(2d,4) \eqa {1\over (2d-1)(2d+2)}\left({(2d)!\over 
  [d!]^2}\right)^2\\[2mm]  
(vi)\quad Dof(2d,5) \eqa {3\cdot 2^{d-2}\over 
(2d-1)(2d+4)(2d+1)^2}\left({(2d+2)!\over [(d+1)!]^2}\right)^2\\[2mm]  
(vii)\quad  Dof(2d,6) \eqa {12\over (2d-1)(2d+6)(2d+1)^2(2d+4)^2}
\left({(2d+2)!\over   [(d+1)!]^2}\right)^2 . \quad \\[-3mm] \nonumber
\eea 
In particular, in $D=4$ one recognizes the two polarizations
of  massless particles of spin $N/2$.
The cases of $N=3$ and $N=4$ correspond to free gravitino
and graviton, respectively, but this is true only in $D=4$.
In other dimensions one has a different field content 
compatible with conformal invariance.

%%%%%%%%%%%%%%%%%%%%%%%%%%%%%%%%%%%
\section{The case of $N=4$ and the Pashnev--Sorokin model}
%%%%%%%%%%%%%%%%%%%%%%%%%%%%%%%%%

For $N=4$ the gauge group is $SO(4)=SU(2) \times SU(2) $.  
Pashnev and Sorokin in \cite{Pashnev:1990cf} considered
the model with a factor $SU(2)$ gauged and the 
other $SU(2)$ left as a global symmetry.
In the analysis of Pashnev and Sorokin the model corresponds to 
a conformal gravitational multiplet, and it was left undecided 
if the field content in $D=4$ is that of a graviton plus three scalars
(five degrees of freedom) 
or that of a graviton plus two scalars (four degrees of freedom).
Thus, we apply the previous techniques to compute the number of 
physical degrees of freedom to clarify the field content of the 
Pashnev--Sorokin model.
As discussed in appendix \ref{sec:PS} the number of degrees of freedom 
of the Pashnev--Sorokin model is given by  
\bea  
Dof(D,{\rm PS}) \eqa  
\frac{1}{2} \int_0^{2\pi} \frac{d\theta}{2\pi} \,    
\Big (2 \cos\frac{\theta}{2} \Big )^{2(D-2)}  
\Big (2 \sin \theta \Big )^2 \ .  \quad
\label{eq:N=2,2}
\eea  
This can be cast in a form similar to those obtained in
section~\ref{section:orto}, and computed explicitly
\bea
Dof(D, {\rm PS}) &=& {2^{2D}\over 2\pi}
\int_0^1 dx\ (1-x)^{D-{3/2}} x^{1/2}= 2^{D-1} {(2D-3)!!\over D!}
\eea
producing $Dof(D,{\rm PS}) = (1,2,5,14,42,132,429,\dots)$
for $D=(2,3,4,5,6,7,8,\dots)$. Thus in $D=4$  one gets 
5 degrees of freedom, which must correspond to a 
graviton plus three scalars. Notice that the 
Pashnev--Sorokin model contains physical degrees of freedom
also in spacetimes of odd dimensions.
Possible couplings of this model to curved backgrounds have been studied in
\cite{Donets:2000sx}.

%%%%%%%%%%%%%%%%%%%%%%% 
\section{Conclusions}
%%%%%%%%%%%%%%%%%%%%%%

We have studied the one-loop quantization of spinning particles
with a gauged $SO(N)$ extended supergravity on the worldline.
These particles describe in first quantization 
all free conformal field equations in arbitrary dimensions and, in particular,
massless fields of higher spin in $D=4$.

We have considered propagation on a flat target spacetime
and obtained the measure on the moduli space of the 
$SO(N)$ supergravity on the circle.
We have used it to compute the propagating physical degrees of freedom
described by the spinning particles.
These models can be coupled to de Sitter or anti de Sitter  backgrounds,
and it would be interesting to study their one-loop partition function
on such spaces. Also,  it would be interesting to study from
the worldline point of view how one could introduce more general couplings, 
giving a different perspective on the problem of constructing
consistent interactions for higher spin fields.

%%%% acknowledgments
\acknowledgments{OC would like to thank the organizers of the
Simons Workshop in Mathematics and Physics 2006, held at SUNY at Stony Brook,
for hospitality and partial support while parts of this work were
completed. He is also grateful to Jac Verbaarschot for a fruitful
conversation concerning orthogonal polynomials.}

\vfill\eject

%%%%%%%%%%%%%%%%%%%%%%%%%%%%%%%%%%%%%%%%%%%%%%%%%%%%%%%%%%%%%
\begin{appendix}
%%%%%%%%%%%%%%%%%%%%%%%%%%%%%%%%%%%%%%%%%%%%%%%%%%%%%%%%%%%%%

%%%%%%%
\section{Gauge fixing of the group $SO(N)$}
\label{gf} 
%%%%%%%

Let us review how the $SO(N)$ gauge fields on the torus 
can be gauge fixed to a set of constant angles taking values 
on the Cartan torus of the Lie algebra of $SO(N)$.
We parametrize the one-dimensional torus of the worldline 
by $\tau \in [0,1]$ with periodic boundary conditions on $\tau$.

Let us start with the simpler $SO(2)=U(1)$ group.
For this case the finite version of the gauge transformations 
(\ref{gauge-tr-e}) looks similar to the infinitesimal one
\bea
a' \eqa a + \dot \alpha \ccr
\eqa  a + \frac{1}{i} g^{-1}\dot g  \ ,  
\qquad g=e^{i \alpha} \in U(1) \ .
\eea
One could try to fix the gauge field to zero by solving
\be
a + \dot \alpha=0  \qquad \Rightarrow \qquad
\alpha(\tau)=-\int_0^\tau dt\,  a(t) \ ,
\ee
but this would not be correct as the gauge transformation 
\be
\tilde g(\tau) \equiv e^{-i \int_0^\tau dt\,  a(t)}
\ee
is not periodic on the torus, $\tilde g(0)\neq \tilde g(1)$. 
In general this gauge transformation is not admissible as it modifies the 
boundary conditions of the fermions. Thus one introduces the constant 
\be
\theta = \int_0^1 dt\, a(t) 
\ee
and uses it to construct a periodic gauge transformation 
connected to the identity (``small'' gauge transformation)
\be
g(\tau) \equiv e^{-i \int_0^\tau dt\,  a(t)}\, e^{i \theta \tau}  \ .
\ee
This transformation brings the gauge field to a constant value on the torus
\be
a'(\tau)  = \theta \ .
\ee
Now ``large'' gauge transformations $e^{i \alpha(\tau)}$
with $\alpha(\tau) = 2 \pi n \tau$ are periodic and
allow to identify 
\be
\theta \sim \theta+ 2 \pi n \ , \quad \quad n\ {\rm integer}\ .
\ee
Therefore $\theta$ is an angle, and one can take $\theta \in [0,2\pi]$ 
as the fundamental region of the moduli space for the $SO(2)$ gauge fields 
on the one-dimensional torus. 

The general case of $SO(N)$ can be treated similarly, using
path ordering prescriptions to take into account the 
non-commutative character of the group.
Finite gauge transformations can be written as
\bea
a' \eqa  g^{-1}a g + \frac{1}{i} g^{-1}\dot g \ ,  
\qquad g=e^{i \alpha}  \ , \qquad \alpha\in {\rm Lie}(SO(N))  \ .
\eea
One can define the gauge transformation
\be
\tilde g(\tau) = {\rm P} e^{-i \int_0^\tau dt\,  a(t)}
\ee
where ``P'' stands for path ordering. 
This path ordered expression solves the equation
\be
\partial_\tau \tilde g(\tau) = -i a(\tau) \tilde g(\tau) 
\ee
and could be used to set $a'$ to zero,
but it is not periodic on the torus, $\tilde g(0)\neq \tilde g(1)$,
and thus is not admissible.
Therefore one identifies the Lie algebra valued constant $A$ by
\be
e^{-i A } = {\rm P} e^{-i \int_0^1 dt\, a(t)}
\ee
so that the gauge transformation given by
\be
g(\tau) \equiv {\rm P} e^{-i \int_0^\tau dt\, a(t)}\, e^{i A \tau } 
\ee
is periodic and brings the gauge potential equal to a constant
\bea
a'(\tau)  \eqa  A \ .
\eea
Since the constant $A$ is Lie algebra valued, 
it is given in the vector representation 
by an antisymmetric $N\times N$ matrix, which can 
always be skew diagonalized by an orthogonal transformation to produce
eq. (\ref{ngauge1}) or eq. (\ref{gauge2}),
depending on whether $N$ is even or odd.
One can recognize that the parameters $\theta_i$ contained in
the latter equations are angles, since one can 
use ``large'' $U(1)$ gauge transformation 
contained in $SO(N)$ to identify
\be
\theta_i \sim \theta_i+ 2 \pi n_i \ , \quad \quad n_i\ {\rm integer}\ .
\ee
The range of these angles can be taken as
$\theta_i \in [0,2\pi]$ for $i=1,\dots,r$, 
with $r$ the rank of the group.
Further identifications restricting the range to a fundamental region
are discussed in the main text.
\vfill\eject

%%%%%%% 
\section{The Van der Monde determinant and orthogonal polynomials} 
\label{sec:VdM}
%%%%%%%
 
In this appendix we briefly review some properties of the Van
der Monde determinant and the orthogonal polynomials method. 
Further details and applications of the method can be found in 
Mehta's book on random matrices~\cite{mehta}.

The Van der Monde determinant is defined by 
\bea 
\Delta (x_i) = \prod_{1\leq k< l\leq r} (x_l-x_k) =  
\left| 
\begin{array}{ccc} 
x_1{}^0 & \cdots & x_r{}^0\\ 
x_1{}^1 & \cdots & x_r{}^1\\ 
: &  & :\\[-3mm] 
\cdot &  & \cdot\\ 
x_1{}^{r-1}& \cdots & x_r{}^{r-1} 
\end{array} 
\right| 
\eea 
where the second identity can be easily proved by induction, observing that: 
(i) the determinant on the right hand side 
vanishes if $x_r=x_i,\ i=1,\dots,r-1$, and 
(ii) the coefficient of $x_r{}^{r-1}$ is the determinant of order $r-1$. 
Furthermore, using basic theorems of linear algebra the Van der Monde 
determinant can be written as 
\bea 
\Delta (x_i) =  
\left| 
\begin{array}{ccc} 
p_0(x_1) & \cdots & p_0(x_r)\\ 
p_1(x_1) & \cdots & p_1(x_r)\\ 
: &  & :\\[-3mm] 
\cdot &  & \cdot\\ 
p_{r-1}(x_1)& \cdots & p_{r-1}(x_r) 
\end{array} 
\right| 
\eea 
where $p_k(x)$ is an arbitrary, order$-k$ polynomial in the variable $x$, 
with the only  constraint of being {\em monic}, that is 
$p_k(x)= x^k+a_{k-1} x^{k-1}+\cdots$.   
 
Interesting properties are associated with the square of the Van der Monde
determinant, which can be written as 
\bea 
\Delta^2(x_i)
\eqa
\det 
\left (
\begin{array}{ccc} 
p_0(x_1) & \cdots & p_{r-1}(x_1)\\ 
p_0(x_2) & \cdots & p_{r-1}(x_2)\\ 
: &  & :\\[-3mm] 
\cdot &  & \cdot\\ 
p_0(x_r)& \cdots & p_{r-1}(x_r) 
\end{array} 
\right ) 
\left (
\begin{array}{ccc} 
p_0(x_1) & \cdots & p_0(x_r)\\ 
p_1(x_1) & \cdots & p_1(x_r)\\ 
: &  & :\\[-3mm] 
\cdot &  & \cdot\\ 
p_{r-1}(x_1)& \cdots & p_{r-1}(x_r) 
\end{array} 
\right )  
\ccr[2mm]
\eqa
\det K(x_i,x_j) 
\eea 
where the kernel matrix $K$ reads as 
\bea 
K(x_i,x_j) = \sum_{k=0}^{r-1} p_k(x_i) p_k(x_j)~. 
\eea 
The above polynomials can be chosen to be orthogonal with respect 
to a certain positive weight $w(x)$ in a domain $D$ 
\bea 
\int_{D} dx\ w(x) p_n(x) p_m(x) = h_n\, \delta_{nm}~. 
\eea 
However, monic polynomials cannot in general be chosen to be 
ortho{\em normal}. Of course, one can 
relate them to a set of orthonormal polynomials $\tilde p_n(x)$ 
\bea 
p_n(x) = \sqrt{h_n}\, \tilde p_n(x) 
\eea     
and the square of the Van der Monde determinant
can be written in terms of a rescaled kernel 
\bea 
\Delta^2 (x_i) = \prod_{k=0}^{r-1} h_k \det \tilde K(x_i,x_j) 
\eea 
with an obvious definition of the latter kernel in terms of the orthonormal 
polynomials. Thanks to the orthonormality condition, the 
rescaled kernel can be shown to satisfy the property  
\bea 
\int_D dz\ w(z) \tilde K(x,z) \tilde K(z,y) = \tilde K(x,y)~, 
\eea 
that can be applied to prove (once again by 
induction) the following identity 
\bea 
&&\int_D dx_r\ w(x_r)\int_D dx_{r-1}\ w(x_{r-1})\cdots\int_D dx_{h+1}\ 
w(x_{h+1})\, \det \tilde K(x_i,x_j) 
\ccr
&&\hskip 2cm
= (r-h)! \det \tilde K^{(h)}(x_i,x_j) 
\nonumber 
\eea 
where $\tilde K^{(h)}(x_i,x_j)$ is the order$-h$ minor obtained by removing  
from the kernel the last $r-h$ rows and columns. In particular 
\bea 
&&\int_D dx_r\ w(x_r)\cdots\int_D dx_{1}\ 
w(x_{1})\, \det \tilde K(x_i,x_j) 
\ccr
&&\hskip 2cm
= (r-1)!\int_D dx_1\ w(x_1)  
\tilde K(x_1,x_1) = r! 
\eea    
and 
\bea 
{1\over r!}\int_D dx_r\ w(x_r)\cdots\int_D dx_{1}\ 
w(x_{1})\, \Delta^2(x_i) = \prod_{k=0}^{r-1} h_k~. 
\eea 
\vfill\eject

%%%%%%%
\section{Gauge fixing of the Pashnev--Sorokin model}
\label{sec:PS} 
%%%%%%%

To derive formula (\ref{eq:N=2,2}) for the physical degrees 
of freedom of the Pashnev--Sorokin model we take the action 
(\ref{action:spinning}) and consider the gauging of
a single $SU(2)$ factor of the $SO(4)=SU(2) \times SU(2)$ symmetry group. 
In order to do that let us consider the change of variables
\bea 
\psi^i =\psi^{\alpha\dot\alpha}\left(\sigma^i\right)_{\alpha\dot\alpha} 
\label{transf} 
\eea 
where 
\bea 
\left(\bar\sigma^i\right)^{\dot\alpha \alpha} = (-i1, {\bf 
  \sigma})^{\dot\alpha \alpha} \ , \quad\quad
\left(\sigma^i\right)_{\alpha\dot\alpha} = (i1, {\bf 
  \sigma})_{\alpha\dot\alpha}=-\epsilon_{\alpha\beta} 
\epsilon_{\dot\alpha\dot\beta}\left(\bar\sigma^i\right)^{\dot\beta 
  \beta}~.
\label{sigmas}
\eea 
The transformation (\ref{transf}) can be inverted as~\footnote{Here we make
  use of the well-known properties $\left(\sigma^i \bar\sigma^j+ \sigma^j
  \bar\sigma^i\right)_\alpha{}^\beta = 2 \delta^{ij}\delta_\alpha{}^\beta,\
  \left(\bar\sigma^i\sigma^j+
  \bar\sigma^j\sigma^i\right)^{\dot\alpha}{}_{\dot\beta} = 2 
\delta^{ij}\delta^{\dot\alpha}{}_{\dot\beta},\
\left(\sigma^i\right)_{\alpha\dot\alpha}\left(\bar\sigma_i\right)^{\dot\beta 
  \beta} = 2 \delta_\alpha{}^\beta \delta_{\dot\alpha}{}^{\dot\beta}$.}
\bea 
\psi^{\alpha\dot\alpha} 
=\frac{1}{2}\psi ^i \left(\bar\sigma_i\right)^{\dot\alpha \alpha}~.
\eea 
The reality condition on $\psi^i$, along with the expressions (\ref{sigmas}),
allows to write it also in the form 
\be
\psi^i = \bar\psi_{\alpha\dot\alpha} \left(\bar\sigma_i\right)^{\dot\alpha 
  \alpha}
\ee
with 
\bea 
\bar\psi_{\alpha\dot\alpha}=- \epsilon_{\alpha\beta} 
\, \epsilon_{\dot\alpha\dot\beta} \psi^{\beta\dot\beta} \ .
\eea 
Thus, the fermion part of the lagrangian can be written as
\bea 
\frac{1}{2}\psi^i(\delta_{ij}\partial_\tau-a_{ij})\psi^j =  
\bar \psi_{\alpha\dot\alpha}\left(\delta^\alpha{}_\beta 
\delta^{\dot\alpha}{}_{\dot\beta}\partial_\tau 
-A^\alpha{}_\beta{}^{\dot\alpha}{}_{\dot\beta} \right) 
\psi^{\beta\dot\beta} 
\eea 
where 
\bea 
A^\alpha{}_\beta{}^{\dot\alpha}{}_{\dot\beta} 
=\frac{1}{2} a_{ij}\left(\bar\sigma^i\right)^{\dot\alpha \alpha} 
\left(\sigma^j\right)_{\beta\dot\beta} 
\eea 
and 
\bea 
a_{ij}=\frac{1}{2} \left(\sigma_i\right)_{\alpha \dot\alpha} 
\left(\bar\sigma_j\right)^{\dot\beta \beta} 
A^\alpha{}_\beta{}^{\dot\alpha}{}_{\dot\beta}  \ .
\eea 
The $SU(2)\times SU(2)$ gauge invariance of the action is now manifest. 
To gauge only a $SU(2)$ subgroup one may choose 
\bea 
A^\alpha{}_\beta{}^{\dot\alpha}{}_{\dot\beta}= 
\delta^\alpha{}_\beta{} B^{\dot\alpha}{}_{\dot\beta}
\quad
\Rightarrow 
\quad
a_{ij} = \frac{1}{2}{\rm tr}\left( \sigma_i B \bar\sigma_j\right) 
\eea 
and gauge fix $B$ to  
\bea 
B^{\dot\alpha}{}_{\dot\beta} 
= 2\theta\, (\frac{i}{2} \sigma^3)^{\dot\alpha}{}_{\dot\beta} 
= i\theta\, (\sigma^3)^{\dot\alpha}{}_{\dot\beta} 
\label{c10}
\eea 
which gives 
\bea 
a_{ij}=\theta\left (
\begin{array}{cccc} 
0&0&0& -1\\ 
0&0&-1&0\\ 
0&1&0&0\\ 
1&0&0&0 
\end{array}\right)
\eea 
so that 
\bea 
\int D\psi\ \exp\left(-{1\over 2}\int 
\psi^i_\mu(\partial_\tau\delta_{ij}-a_{ij})\psi^j_\mu\right)  
&=& {\rm Det}^D(\partial_\tau+i\theta)_{_{ABC}}\ 
{\rm Det}^D(\partial_\tau-i\theta)_{_{ABC}} 
\ccr 
&=& \left(2\cos{\theta\over 2}\right)^{2D}~. 
\eea  
The Faddeev-Popov determinant associated to the gauge-fixing 
of the $SU(2)$ gauge group reads 
\bea 
{\rm Det}\, (\partial_\tau 1_{adj}-B_{adj})_{_{PBC}}=   
\left(2\sin \theta\right)^{2} 
\eea 
since eq. (\ref{c10}) in the adjoint representation becomes
\bea 
B_{adj}=2\theta\left( 
\begin{array}{ccc} 
0&-1&0\\ 
1&0&0\\ 
0&0&0 
\end{array}\right)~. 
\eea 
Finally, the Faddeev-Popov determinant associated to gauge-fixing the 
local supersymmetry reads 
\bea 
{\rm Det}^{-1}(\partial_\tau\delta_{ij}-a_{ij})_{_{ABC}}= 
\left(2\cos{\theta\over 2}\right)^{-4}  \ .
\eea 
Assembling all determinants one gets (\ref{eq:N=2,2}), 
where the factor $1/2$ is due to the parity transformation
$\theta \to - \theta$.
\end{appendix} 
\vfill\eject

%%%%%%%%%%%%%%%%%%%%%%%%%%%%%%%%%%%%%%%%%%%%%%%%%%%%%%%%
%%%%%%%%%%%%%%%%%%%%%%%%%%%%%%%%%%%%%%%%%%%%%%%%%%%%%%%%%% 

\end{document}